\documentclass[%
 aip,
 amsmath,
 amssymb,
 reprint,%
]{revtex4-1}

\usepackage[utf8]{inputenc}
\usepackage{graphicx}  
\usepackage{dcolumn}   
\usepackage{bm}        

\usepackage[T1]{fontenc}

\usepackage{mathptmx}
\usepackage{etoolbox}

\usepackage[euler]{textgreek}
\usepackage{upgreek}
\usepackage[american]{babel}
\usepackage{xcolor}
\usepackage{comment}
\usepackage{ulem}

\makeatletter
\def\@email#1#2{%
 \endgroup
 \patchcmd{\titleblock@produce}
  {\frontmatter@RRAPformat}
  {\frontmatter@RRAPformat{\produce@RRAP{*#1\href{mailto:#2}{#2}}}\frontmatter@RRAPformat}
  {Guido Dittrich}{}
}%
\makeatother

\begin{document}

\preprint{AIP/123-QED}

\title{Polymeric Liquids in Nanoporous Photonic Structures: From Precursor Film Spreading to Imbibition Dynamics at the Nanoscale}
\author{Guido Dittrich}
\affiliation{Institute for Materials and X-Ray Physics, Hamburg University of Technology, 21073 Hamburg-Harburg, Germany}

\author{Luisa G. Cencha}
\affiliation{IFIS-Litoral (Universidad Nacional del Litoral-CONICET), Guemes 3450, 3000 Santa Fe, Argentina}

\author{Martin Steinhart}
\affiliation{Institute of Chemistry of New Materials, Osnabrück University,49076 Osnabrück, Germany}

\author{Ralf B. Wehrspohn}
\affiliation{Institute of Physics, Martin Luther University of Halle-Wittenberg, 06120 Halle (Saale), Germany}
\affiliation{Korea Institute of Energy Technology (KENTECH), 21 Kentech-gil, Naju 58330, South Korea}

\author{Claudio L. A. Berli}
\affiliation{INTEC (Universidad Nacional del Litoral-CONICET), Predio CCT CONICET Santa Fe, RN 168, 3000 Santa Fe, Argentina}

\author{Raul Urteaga}
\affiliation{IFIS-Litoral (Universidad Nacional del Litoral-CONICET), Guemes 3450, 3000 Santa Fe, Argentina}

\author{Patrick Huber}
\affiliation{Institute for Materials and X-Ray Physics, Hamburg University of Technology, 21073 Hamburg-Harburg, Germany}
\affiliation{Center for X-Ray and Nano Science CXNS, Deutsches Elektronen-Synchrotron DESY, 22603 Hamburg, Germany}

\date{\today}

\begin{abstract}
Polymers are known to wet nanopores with high surface energy through an atomically thin precursor film followed by slower capillary filling. We present here light interference spectroscopy using a nanoporous membrane-based chip that allows us to observe the dynamics of these phenomena in situ with sub-nanometer spatial and milli- to microsecond temporal resolution. The device consists of a mesoporous silicon film (average pore size 6 nm) with an integrated photonic crystal, which permits to simultaneously measure the phase shift of the thin-film interference and the resonance of the photonic crystal upon imbibition. For a styrene dimer, we find a flat fluid front without a precursor film, while the pentamer forms an expanding molecular thin film moving in front of the menisci of the capillary filling. These different behaviors are attributed to a significantly faster pore-surface diffusion compared to the imbibition dynamics for the pentamer and vice versa for the dimer. In addition, both oligomers exhibit anomalously slow imbibition dynamics, which could be explained by apparent viscosities of six and eleven times the bulk value, respectively. However, a more consistent description of the dynamics is achieved by a constriction model that emphasizes the increasing importance of local undulations in the pore radius with the molecular size and includes a sub-nanometer hydrodynamic dead, immobile zone at the pore wall, but otherwise uses bulk fluid parameters. Overall, our study illustrates that interferometric, opto-fluidic experiments with nanoporous media allow for a remarkably detailed exploration of the nano-rheology of polymeric liquids. 
\end{abstract}

\maketitle

\maketitle

\section{Introduction}

Polymeric liquids exhibit a variety of transport mechanisms during capillary filling of nanoporous materials. Understanding them is essential for many applications. Immobile layers at the pore walls or confinement effects are important for separation processes\cite{Krutyeva2013, Kusmin2010}, the formation of precursor films can be used for templated tubes\cite{Steinhart2002} and temperature-dependent imbibition for passive temperature sensors \cite{Cencha2022}, to just name a few.\\

Discussions describing capillary rise in nanopores often start with the Lucas-Washburn equation (LWE).\cite{Lucas1918,Washburn1921} It relates the driving capillary pressure to the counteracting viscous drag in a ${\rm \sqrt{t}}$ versus imbibition length ${\rm L}$ dependency. Although it is not suitable to solve every capillary filling problem, in particular if there are systematic variations in the hydraulic permeability, \cite{Reyssat2008} it is fairly close to many solutions of more complex porous systems and liquids.\cite{Gruener2009,Gruener2015, Huber2015} The prefactor of the LWE can be divided into the fluid-centered properties; surface tension ${\rm \sigma}$, viscosity ${\rm \mu}$, and the liquid-solid contact angle ${\rm \theta}$, and into the geometrical ones; an effective pore radius ${\rm r_{\rm eff}}$ whose definition captures the nature of the porous structure.

\begin{eqnarray}
   {\rm L(t)=\Gamma \sqrt{r_{\rm eff}t}	 }
    \label{eq:LW}
\end{eqnarray}

where $\Gamma=\sqrt{\frac{\sigma \cos{\theta}}{2 \mu}}$.

Many experimental studies have been performed on melt infiltration of polymers, with a molecular weight above the critical value (${\rm M_c}$) for entanglements, into nanopores.\cite{VazquezLuna2021a,Luna2022,Cencha2019,yao_complex_2017}
In such studies\cite{VazquezLuna2021a,Luna2022} on melt infiltration of polystyrene into anodized aluminum oxide with straight pores and into the sponge-like structure of controlled porous glass, a square-root-of-time law was used to model the dynamics. Pearson correlation coefficients for different exponents of a power-law fit suggested though an even better fit for exponents deviating from the famous LWE like ${\rm \sqrt{t}}$ proportionality.\\ 

Yao et al. \cite{yao_theory_2018} proposed a unified theory of capillary rise dynamics of polymers ${\rm >M_c}$ involving an immobile layer and reptation under pressure model. The latter explains the relative increase of imbibition dynamics with molecular mass, in contrary to the increasing bulk viscosity, by disentanglement under confinement. In the immobile layer model the pore radius behind the imbibition front is reduced by adsorbed molecules. This phenomenon is also observed for polymeric liquids with molecular weight below ${\rm M_c}$ and other liquids.\\

Polymeric liquids below ${\rm M_c}$ under confinement exhibit imbibition phenomena that are not well understood yet. A molecular dynamics simulation of Lennard-Jones (LJ) fluids and a decane melt in a cylindrical nanopore is in close agreement with the LWE.\cite{dimitrov_capillary_2007} In case of slip flow they suggest to modify the LWE by virtually increasing the radius by a slip length. Also a much faster precursor film spreading obeying a square-root-of-time law has been inferred in atomistic simulations \cite{Chibbaro2008a}.\\  

Engel and Stühn\cite{Engel2010} studied polyisobutylene and poly-${\rm \epsilon}$-caprolactone in nanoporous polycarbonate (both <${\rm M_c}$) by in-situ small angle X-ray scattering. They found a fast first wetting of the pore walls by a precursor film, followed by a much slower than expected complete filling. The fast initial wetting is attributed to a stronger adhesive than cohesive force. Although the aspect ratio of pore size to the radius of gyration was not small at all, they discovered a strong confinement effect on the complete filling.\\ 
In-situ nanodielectric spectroscopy is simultaneously sensitive to molecular dynamics and capillary rise. An imbibition study\cite{tu_interfacial_2020} on cis-polyisoprene (<${\rm M_c}$) in anodized aluminum oxide (AAO) has measured two time regimes decelerated compared to the LWE. An increasing molecule-wall interaction with time for native AAO was concluded, as less freely fluctuating chains were contributing to the dielectric signal. Modifying the surface by silanization successfully reduced the interaction.\\

Recent single-pore molecular dynamics simulations of LJ fluids predict slower capillary filling for short polymer chains, below ${\rm M_c}$ for entanglements, compared to the LWE prediction, while longer chains show the opposite trend.\cite{zhang_capillary_2023} The slowed down filling dynamics of short polymer chains is explained by a nearly immobile layer with finite slip and lower free energy due to confinement. It is quantified by an effective viscosity, which is twice the bulk viscosity for a radius-to-chain length ratio of one and, interestingly, increases for higher ratios.\\

We focus here on the imbibition dynamics of styrene oligomers in nanoporous silicon and will see that our observation fit such a slowdown seen in the simulations. Moreover, we are going to scrutinize the relevance of precursor film spreading for the imbibition phenomenology.\\    

Nanoporous silicon (pSi) is an established functional material \cite{sailor2012porous, Canham2015, Brinker2020, Brinker2022,Brinker2022a}, most prominently in optical sensing applications for biochemistry\cite{pacholski_photonic_2013}, gases\cite{Lin2004AHydrogen} and fluid dynamics\cite{cencha_precursor_2020}. It can also be used to explore the effect of nanoconfinement on condensed matter. \cite{Henschel2007,Hofmann2012,Calus2012,Huber2015, Kondrashova2017} A common optical sensing mechanism relies on the change of the effective refractive index of porous layers during the displacement of air by an analyte. Porous silicon is electrochemically etched in a self-organizing top-down process with a strong anisotropy of the pores in the etching direction. When the pore size and interpore distances are much smaller than the probing wavelength of an electromagnetic wave, the effective optical properties of pSi are adjustable by the porosity and thickness. The latter parameters depend on the synthesis conditions. Continuous or abrupt changes in the in-depth porosity can be accomplished by the course of applied current density, whereas the time applied determines the thickness. This enables the synthesis of layered structures, Bragg- and Rugate-filters, as well as 1-D photonic crystals.\cite{pacholski_photonic_2013} These  structures can be applied for the in-situ investigation of polymer infiltration into nanoporous scaffold materials,\cite{cencha_precursor_2020} providing an improved resolution in the plane perpendicular to the pore orientation.

\section{problem statement}
It has been demonstrated that in-situ white light reflection spectroscopy of pSi photonic crystals (PC) provides the precision to resolve precursor film spreading. A thin film pSi layer in front of the PC (\ref{fig:setup_constricted}a,b) acts as a column to investigate different imbibition lengths and thereby times. The shape of fluid fronts can be inferred from re-scaling the filling dynamics of the PC with a capillary-filling model for the fluid transport of the in-front homogeneous porous layer.\cite{cencha_precursor_2020}\\

As outlined above, polymers exhibit a variety of different dynamics for the infiltration of nanoporous scaffolds \cite{yao_theory_2018}, depending on the molecular weight, temperature, entanglements, shear thinning/thickening\cite{cao_capillary_2016} behavior or confinement. Therefore, the capillary filling model that best fits the experimental data needs to be adapted for each individual fluid. The geometrical descriptor of the scaffold imbibed should on the contrary be unanimous. \\

In the following a method is introduced, where the infiltration of the stacked pSi-layer (thin film + PC) is measured simultaneously with two spectrometers comprising two different wavelength ranges; visible and NIR. In this way, the infiltration curve corresponding to the PC can be decoupled from the infiltration curve of the stacked layer. Both curves, combined, allow us to infer not only the meniscus dynamics but also the precursor film profile and dynamics. In this study we have used two oligostyrenes as imbitition fluids. Overall, we introduce an exceptionally precise technique to characterize fluid dynamics, particularly relevant for polymer infiltration into nanopores.

\section{materials}

Probing liquids are mono disperse oligostyrenes (OS) (from Polymer Standards Service) with two (OS-2) and five (OS-5) monomer repetition units, having a sec-butyl and proton end group. Because of their well defined properties, OS molecules are commonly used as a standard for gel-chromatography. The viscosity at room temperature is analysed by a rheometer (Kinexus Prime Series) in plate to plate geometry, resulting in a viscosity of 34,000${\rm {\pm}570~mPas }$ in a shear rate range of ${\rm0.01-100~Hz}$ and 9.4${\rm {\pm}2.1~mPas }$ at ${\rm 100-51600~Hz}$ for OS-5 and OS-2 respectively. Both liquids provide Newtonian fluid behavior in the shear-rate range of the capillary filling. Other parameters used in the analysis are listed in Tab.~\ref{tab:parameters}.  

Porous silicon is fabricated by electrochemical anodization. Monochristalline 100-orientation, boron doped (p-type) wafers with resistivity of ${\rm 1-4~m{\Omega}}$cm are treated with a HF(${\rm 50~w\%}$):EtOH 1:2 volumetric solution and different current densities. By variation of the latter, layers with different thicknesses and porosity can be obtained. For the current method, a thin film layer of largely homogeneous porosity is etched first, subsequently followed by a PC. The PC is designed with its band gap and width in the visible range, whereas the overall stack of the layers has a detectable interference pattern in the NIR range of the used setup. The PSi thin films were fabricated using a current density of 12.7 mA $\rm cm^{-2}$, while the PC was fabricated by
alternating current densities of 12.7 and 50.9 mA $\rm cm^{-2}$.  More information about the synthesis can be found in a previous publication \cite{cencha_precursor_2020}.

\begin{table}
\caption{\label{tab:parameters} Literature values used in the calculations. The refractive indices are the
mean between ${\rm 1150-1650~nm}$ for the RIFTS assessment of porosity and thickness.}
\begin{ruledtabular}
\begin{tabular}{cccccc}
 ${\rm \sigma}$\cite{wu1970surface} [${\rm mJ/m^2}$] & ${\rm cos(\theta)}$\cite{xu2022solid} & ${\rm r_h}$ [nm] & ${\rm n_{\rm Si}}\cite{green2008self}$ & ${\rm n_{\rm air}}$\cite{peck1972dispersion} & ${\rm n_{\rm OS}}$\cite{zhang2020complex}\\
\hline
40.7 & 1 & 5.8 & 3.51 & 1 & 1.57
\end{tabular}
\end{ruledtabular}
\end{table}

\section{Setup}

\begin{figure*}
 \includegraphics[width=0.85\textwidth]{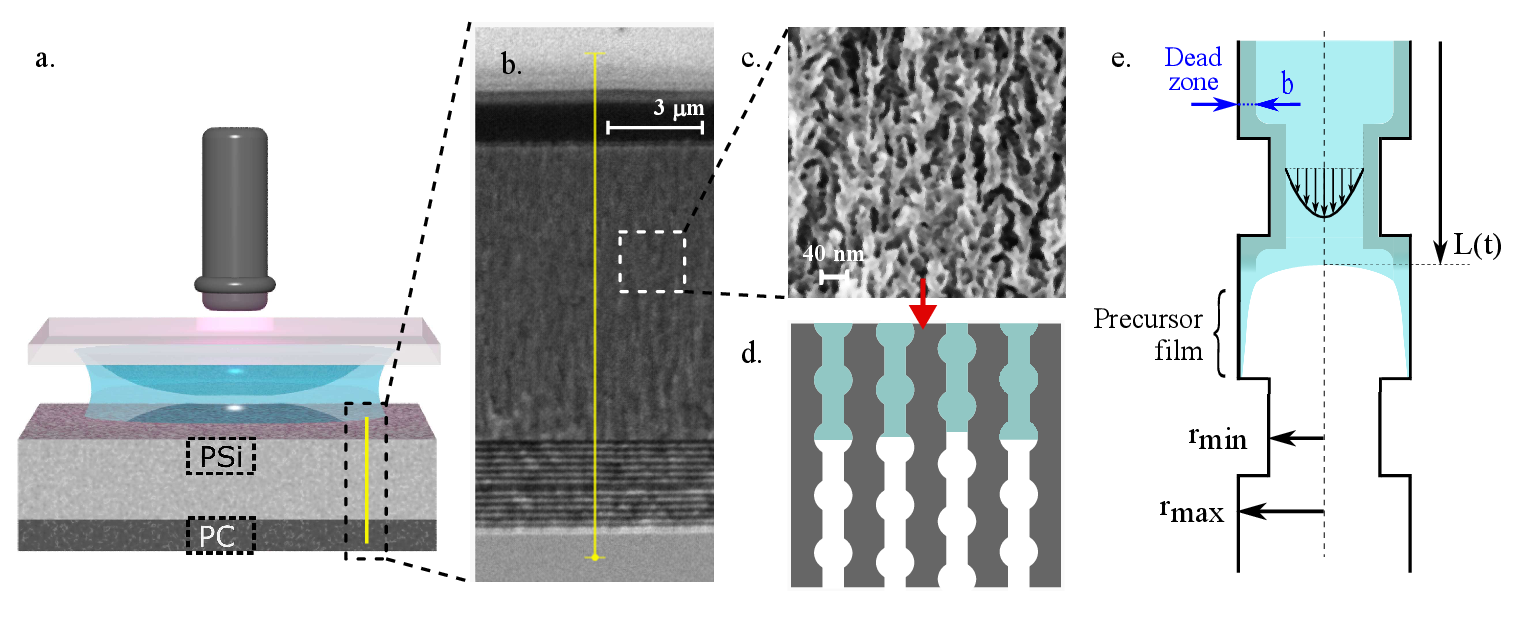}
 \caption{\label{fig:setup_constricted} a. Illustration of the measurement setup, starting from the top consisting of a reflection probe, glass carrier, oligomer film, and sample. b.  Scanning electron micrograph of the sample regions indicated in the illustration. c.  PSi structure with a strong anisotropic pore orientation in the direction of the etching although the displayed pore walls on the given scale appear dendritic. d. Simplified representation by a constriction model. e. Single representative, constricted tube with a dead zone for the fluid velocity profile during spontaneous imbibition.    }
\end{figure*}

Measurements are conducted with a white light reflection spectroscopy setup (Fig.~\ref{fig:setup_constricted}) consisting of a glass fiber bundle reflection probe with 17 illumination and two read fibers. Illumination is provided by a balanced deuterium-halogen light source (AVALIGHT-DH-S-BAL) with a spectral range of ${\rm 215-2500~nm}$. One read fiber is connected to an UV-VIS-spectrometer (AvaSpec-2048CL) equipped with a ${\rm 10~\mu m}$ slit, ${\rm 200-1100~nm}$ spectral range and a spectral resolution of 1~nm. The other to an, via a coax cable synchronized, NIR-spectromter (AVASPEC-NIR512) with a ${\rm25~\mu m}$ slit, ${\rm 850-1650~nm}$ spectral range and a spectral resolution of 4~nm.

\section{Method}

The imbibition of oligostyrenes is monitored by in-situ white light spectroscopy at normal incidence and conducted at room temperature. Before the experiment PSi samples are cleaned with toluene and allowed to dry. The oligomers are drop-cast on a glass substrate and fixed in front of the reflection probe (see Figure \ref{fig:setup_constricted}a). Spontaneous capillary rise is initiated once the polymer film is brought in direct contact with the sample. The sampling rates of the spectrometers where chosen to 500~ms for OS-5 and ${\rm 1-5~ms}$ for the lower viscosity OS-2, respectively.   \\
At first, the PSi thin film is infiltrated. The substitution of air by OS with its higher refractive index increases the effective refractive index ${\rm n_{\rm eff}}$ of the material, which can be measured by a proportional shift of the optical thickness (OT) ${\rm e_{\rm ot}}$ according to ${\rm  e_{\rm ot}=2d{\cdot}n_{\rm eff}}$, where d is the physical thickness. Assuming an endless supply of liquid during the capillary filling, a normalized OT ${\rm {\Delta}e_{\rm ot}(t)}$ can be determined by the values at the starting and completed filling time ${\rm t_0}$ and ${\rm t_{\rm end}}$, respectively. This normalized OT reflects the filling fraction of the imbibition. In the case of homogeneous porosity, which implies a homogeneous $\rm n_{\rm eff}$, through the entire thickness, ${\rm {\Delta}e_{\rm ot}(t)}$ is directly related to the imbibition fluid front position. For other pore morphologies, ${\rm {\Delta}e_{\rm ot}(t)}$ accounts for the relative pore volume infiltrated. With the range of layer thicknesses between ${\rm3-20~\mu m}$, it was found that the OT is determined with the best signal to noise ratio in the wavelength range of ${\rm1150-1650~nm}$.\\ 

\begin{figure}
 \includegraphics[width=0.45\textwidth]{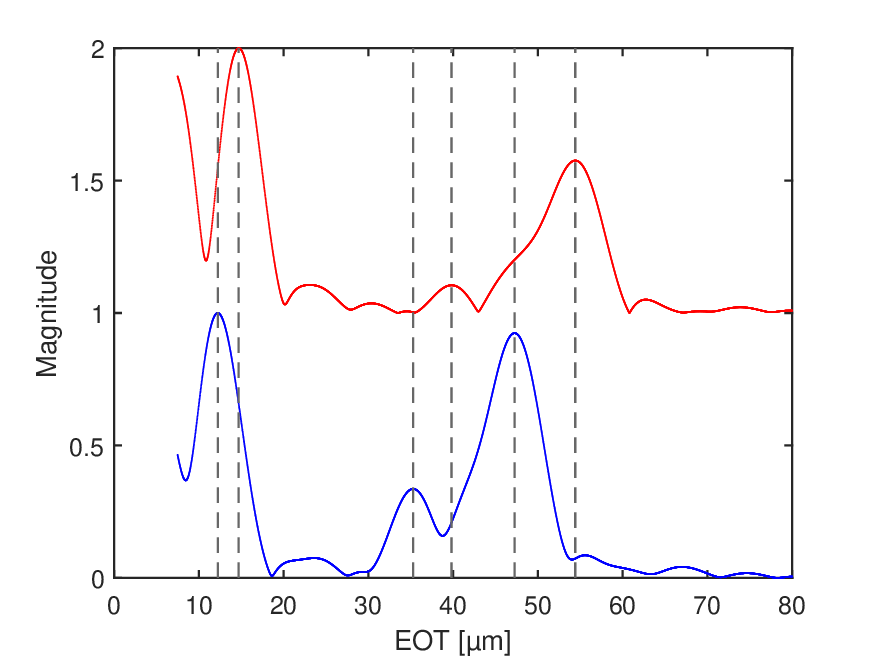}
 \caption{\label{fig:FFT} Fast Fourier transform (FFT) of the initial (blue) and final (red) spectrum in the NIR region from ${\rm 1150-1650~nm}$. For illustration purposes the magnitude of the final curve is shifted by a magnitude of one. The dashed lines indicate the center of the main peaks by full width half maximum (FWHM) peak determination. Start (left) and end (right) are used to solve the orthogonal Maxwell-Garnett effective medium approximation (Eq.~\ref{eq:oMG}) for the layers empty and filled state as described in the method section. Thickness and porosity of the layers are listed in Tab.~\ref{tab:EMA}. }
\end{figure}

For a single thin film a simple sinusoidal interference pattern would be measured that undergoes a phase shift during the capillary filling. This can be tracked by several methods. Because of the stacked structure of PSi thin film and PC, a superposition of sinusoids is measured. A mathematical transformation method facilitates differentiating the sinusoids and attributing them to the respective physical layer. Therefore, a Hann-windowed fast Fourier transformation (FFT) is used for the analysis of the OT. This method is known as reflective interferometric Fourier transform spectroscopy (RIFTS)\cite{sailor2012porous}. Fig.~\ref{fig:FFT} displays the FFT of an empty and subsequently OS-filled sample. The three most prominent peaks, indicated by vertical dotted lines, can be attributed to the PC, the thin film, and the stack of both from left to right. This is conducted by effective medium approximation with the orthogonal Maxwell-Garnett model Eq.~\ref{eq:oMG}\cite{jalas_effective_2014}, where ${\rm  \Phi_i}$ is the average porosity of the layer i, ${\rm  n_{\rm \rm 1/2}}$ is the refractive index of the filling medium and ${\rm  n_{\rm \rm Si }}$ the one of silicon, in conjunction with ${\rm  e_{\rm \rm ot}=2d{\cdot}n_{\rm \rm eff}}$. \\
The results of the thicknesses are in good agreement with scanning electron microscopy and the porosities are consistent (Tab.~\ref{tab:EMA}). An approach to measuring the capillary filling would be to follow the particular OT peak in the FFT over time. This has been found challenging for the PC and thin film peak, as they are not always apparent, since the thin film peak is too close to the peak of the overall stack. Here, a major problem arises from windowing, which increases the width of the peaks, but is also a necessity to reduce spectral leakage and thereby side lobes of the main peaks. The solution chosen is to follow the OT of the stack (see Figure S2). An equivalent alternative technique is to follow the wavelength temporal evolution of a certain peak of the interference pattern in the spectrum, $\rm \lambda_{\rm peak}(t)$. In this method, the resulting data series is normalized by initial and final wavelength. The latter method requires less processing time and because of that is the chosen one in this work. A comparison of the curves obtained by both methods can be found in Figure S2. \\

\begin{eqnarray}
{\rm  {\Delta}e_{\rm ot}(t)=\frac{e_{\rm ot}(t)-e_{\rm ot}(t_0)}{e_{\rm ot}(t_{\rm end})-e_{\rm ot}(t_0)}}
    \label{normalizedEOT}
\end{eqnarray}

\begin{table}
\caption{\label{tab:EMA} Thickness ${\rm d_{\rm oMG}}$ and porosity ${\rm \Phi_{\rm oMG}}$ by orthogonal MG using the RIFTS method\cite{sailor2012porous} for the measurement displayed in Fig.~\ref{fig:FFT}. The thickness is in reasonable agreement with SEM measurements and the equation ${\rm d_{\rm tot} \Phi_{\rm tot}=d_{\rm film} \Phi_{\rm film} + d_{\rm PC} \Phi_{\rm PC}}$ is fulfilled.}
\begin{ruledtabular}
\begin{tabular}{cccc}
layer & ${\rm d_{\rm SEM}}$  [${\rm \mu m}$] & ${\rm d_{\rm oMG}}$ [${\rm \mu m}$] & ${\rm \Phi_{\rm oMG}}$ [${\rm \%}$]\\
\hline
PC & 3.4& 3.5 & 71\\
thin film & 9.3& 8.3 & 55\\
stack & 12.7 & 11.9 & 61\\
\end{tabular}
\end{ruledtabular}
\end{table}

\begin{eqnarray}
{\rm  n_{\rm 1/2}^2=\frac{\Phi_i \left ( n_{\rm 1/2}^2 \frac{2n_{\rm Si}^2}{n_{\rm 1/2}^2+n_{\rm Si}^2} - n_{\rm Si}^2 \right )+n_{\rm Si}^2}{\Phi_i \left (\frac{2n_{\rm Si}^2}{n_{\rm 1/2}^2+n_{\rm Si}^2} -1\right )+1}}
\label{eq:oMG}
\end{eqnarray}

Subsequently the PC is filled by the invading OS. Here the filling of the central cavity can be measured by the increasing wavelength of its resonance valley.\cite{cencha_precursor_2020}  It can be normalized to start and end analogously to the OT in Eq.~\ref{normalizedEOT}.  Abstractly viewed, the PC acts as a sensor of the local filling, since its resonance position ${\rm \lambda_0}$ is extremely sensitive to the optical thickness of its central defect layer. This means that when even a small fraction of air is replaced by an invading liquid, the microcavity will change the original syntonization wavelength showing a shift in ${\rm \lambda_0}$.

\section{analysis}

In the following, capillary rise in pSi is treated as its structure consists of an unconnected array of pores, which can be described by a representative effective pore radius $\rm r_{\rm eff}$. The dynamics of fluids for this kind of problem are often described or at least compared to the LWE. Since many studies of more complex liquids and geometries have discovered deviations from this law, several modified LWE like square-root-of-time laws have been proposed to adapt.\\

  Cencha et al. \cite{cencha_interferometric_2018} have demonstrated that the melt infiltration of an ethyl vinyl acetate copolymer into pSi can be described by taking into account a mean hydraulic radius ${\rm \langle r_{\rm h} \rangle= 2A/P}$, defined by the ratio of pore perimeter P and area of the cross-section A, and a tortuosity parameter ${\rm \tau_{\rm L}}$ (Eq.~\ref{eq:tau_L}). The latter is the elongation factor relating the imbibition length, i.e. the normal of the materials surface plane and the center of the imbibition front, to the actual path of a fluid through a meandering pore. In previous experiments with short chained alkanes a value of 2.6 was determined for this tortuosity in pSi.\cite{acquaroli_capillary_2011} \\ 

\begin{eqnarray}
    {\rm r_{\rm eff}=\frac{\langle r_h \rangle}{\tau_L^2} }
    \label{eq:tau_L}
\end{eqnarray}

In that approach, the concept of tortuosity comes to mathematically explain the abnormally low effective radius $\rm r_{\rm eff}$ that is required to describe the experimental results. Actually, $ \rm r_{\rm eff}$  represents a straight cylindrical pore with the same imbibition dynamics. In the present work, we adopt a different strategy to describe the polymer infiltration in the mesoporous silicon: we start by defining a pore network that better represents the observed physical structure (Figure \ref{fig:setup_constricted}d), and then incorporate corrections that compensate for some nanoscale effects, notably the presence of a stagnant polymer layer on the pore walls. Calculations are made in the framework of the sub-continuum approach \cite{kavokine_2021}, where continuum equations are still valid, although emerging confinement effects are to be considered. 

More precisely, we consider an array of nanotubes with periodic modulation of the cross-section, where the pore radius varies between $\rm r_{\rm min}$ and $ \rm r_{\rm max}$ (Figure \ref{fig:setup_constricted}e). Then, momentum and mass conservation equations are applied to calculate the spontaneous imbibition dynamics. Assuming that modulations randomly repeat along the tube, the effective radius can be written as

\begin{eqnarray}
    {\rm r_{\rm eff} =\frac{1}{\langle r^{-4}\rangle \langle r^{3}\rangle }}
    \label{eq:r_eff_mean}
\end{eqnarray}

where $\rm{\langle r^n\rangle =\frac{1}{h}\int_{\rm 0}^{h} r(z)^n \,dz }$
is the n-th probability moment of the radius distribution, being $h$ the distance across the porous material \cite{Franck2022}. It is worth noting that Eq. \ref{eq:r_eff_mean} had been previously reported, though with a less general formalism \cite{Sharma1991}. In particular, for an array of tubes with periodic step changes of radius, the averages in Eq. \ref{eq:r_eff_mean} are $\langle \rm r^n \rangle = (\rm r^n_{\rm min} + r^n_{\rm max})/2$ and the effective radius results,

\begin{eqnarray}
  \rm{ \frac{r_{\rm eff}}{r_{\rm min}} = \frac{4}{{\left(\frac{r_{\rm min}}{r_{\rm max}}\right)^4 + \left(\frac{r_{\rm max}}{r_{\rm min}}\right)^3 + \frac{r_{\rm min}}{r_{\rm max}} + 1 }}
   \label{eq:r_eff_2}}
\end{eqnarray}

This expression clearly shows that the periodically constricted tube yield an effective radius that can be even lower than that of a uniform cylinder of radius ${\rm r_{ min}}$. Further, for ${\rm r_{max} ⁄ \rm r_{min} \gg 1}$, Eq.~\ref{eq:r_eff_2} can be written as follows,

\begin{eqnarray}
   {\rm\frac  {r_{\rm eff}}{r_{\rm min}} \approx
4 \left(\frac{r_{\rm min}}{r_{\rm max}}\right)^3 }
    \label{eq:r_eff_3}
\end{eqnarray}

The functionality of Eq. \ref{eq:r_eff_3} suitably captures the characteristic length scales governing the capillary filling of mesoporous materials \cite{Berli2017}. In addition, the trend predicted by Eq. \ref{eq:r_eff_3}  has been experimentally observed in different porous media with bimodal pore size distributions \cite{Dullien1977, Patro2007}.

In the case of the pSi thin film used in this work (Fig.~\ref{fig:setup_constricted}), $\rm \langle r_h\rangle=5.8$ nm is obtained from the analysis of SEM images \cite{cencha_precursor_2020}. Moreover, imbibition measurements with glycerol indicate that this structure has $\rm r_{\rm eff}\sim 0.9$ nm, which is consistent with a tortuosity of 2.6 (Eq. \ref{eq:tau_L}).\cite{cencha_precursor_2020} Then, by combining Eqs. \ref{eq:tau_L}, \ref{eq:r_eff_mean} and \ref{eq:r_eff_3}, it is found that the morphology of this layer can be represented considering the model of periodically constricted tubes with $\rm r_{\rm min}=3.5$ nm and $\rm r_{\rm max}=8.5$ nm.

\begin{figure}
\includegraphics[width=0.45\textwidth]{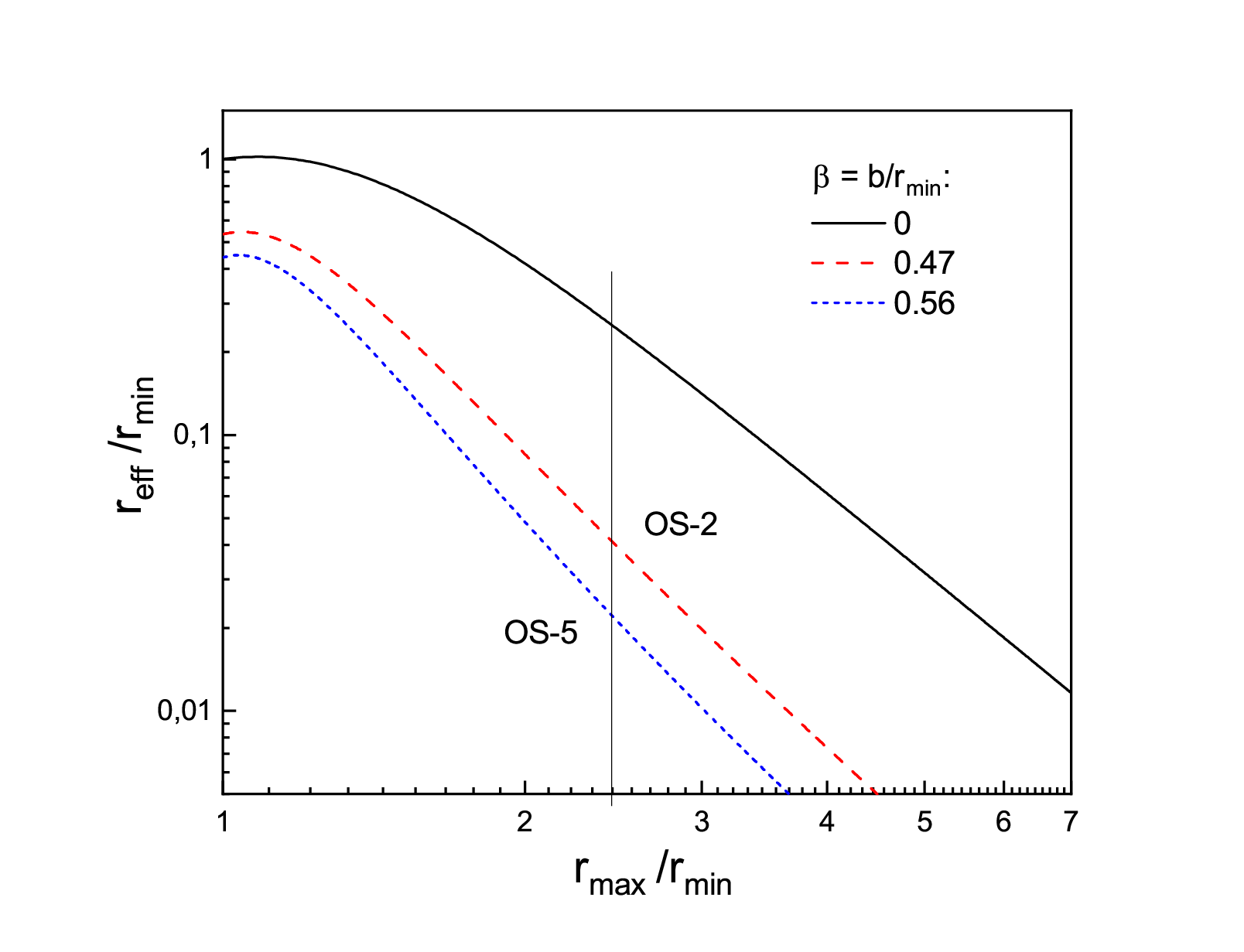}
\caption{\label{fig:Model}  Theoretical prediction of the effective pore radius ($\rm  r_{\rm eff}$) from Lucas-Washburn dynamics in periodically constricted tubes. The continuous line shows the dependence of $\rm r_{\rm eff}$ with the pore size ratio ( $\rm r_{\rm max}⁄r_{\rm min}$, Eq. \ref{eq:r_eff_2}). The dashed lines show the model results when an immobile layer of size $\rm b$ is incorporated (Eq. \ref{eq:r_eff_4}). Results are shown for two values of $\rm b$ that fit the measured behavior of the oligomers OS-2 and OS-5. The vertical full line stands for the pore ratio (2.43) that characterizes the pSi sample used in the experiments. }
\end{figure}

If one further considers the presence of an adsorbed layer of polymer, which forms a dead zone of thickness $b$ in the flow field (Fig.~\ref{fig:setup_constricted}e), then Eq.~\ref{eq:r_eff_2} results 

\begin{eqnarray}
   {\rm \frac{r_{\rm eff}}{r_{\rm min}}=
   \frac{4(1-\beta)}{\alpha^{-4} + \alpha^3 + \alpha^{-1} + 1}}
    \label{eq:r_eff_4}
\end{eqnarray}

where ${\rm \alpha=[(r_{\rm max} ⁄ r_{\rm min})-\beta]⁄(1-\beta)}$ and ${\rm \beta=b ⁄ r_{\rm min}}$ . Figure~\ref{fig:Model} illustrates the prediction of Eq.~\ref{eq:r_eff_4}, where one readily observes the strong diminution of ${\rm r_{\rm eff}}$ induced by a uniform layer of stagnant polymer on the pore walls. 

In what follows we use Eq.~\ref{eq:LW} with this effective radius (Eq. \ref{eq:r_eff_4}) to interpret the capillary filling curves obtained from styrene oligomers.\\

\begin{figure*}
\includegraphics[width=1\textwidth]{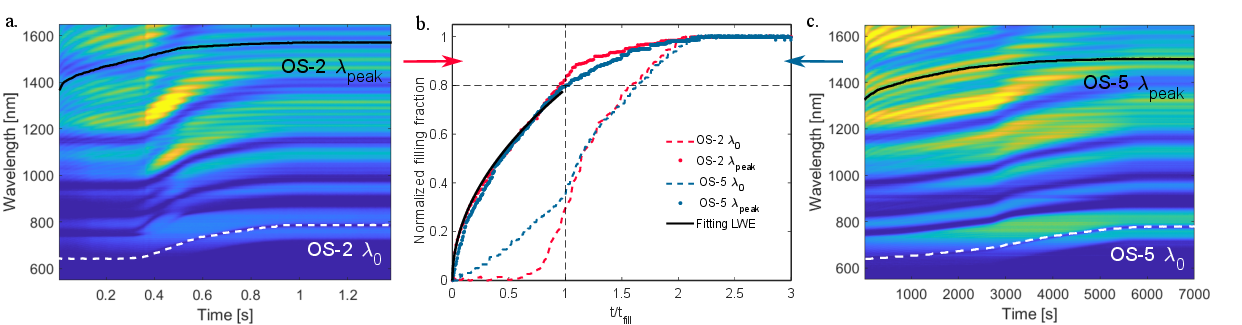}
\caption{\label{fig:Reff_fit} a,c. Contour plots of the capillary filling of pSi samples for OS-2 and OS-5, respectively. The color scale represents the reflectance intensity. Solid lines show the evolution of one reflectance peak $\rm \lambda_{\rm peak}$ during imbibition while dashed lines indicate the evolution of the microcavity position $\rm \lambda_0$. b. Normalized evolution of  $\rm \lambda_{\rm peak}$ and $\rm \lambda_0$ as a function of time normalized with $\rm t_{\rm fill}$. The solid line shows a fitting with a $\rm \sqrt{t}$ dependence.}
\end{figure*}

Figure \ref{fig:Reff_fit} shows the results obtained for the capillary imbibition of the oligomers.  Figure  \ref{fig:Reff_fit}c show a contour map of the evolution of the spectral reflectance for OS-5. A superimposed dashed line indicates the evolution of the resonance wavelength of the cavity formed by the PC. The first increase in wavelength begins almost immediately and is associated with the arrival of the fluid precursor film to the PC. The second abrupt increase that occurs around 3100s is associated with the arrival of the fluid front to the PC. Figure \ref{fig:Reff_fit}b shows in dashed lines the evolution of the microcavity normalized position for measurements carried out with OS-2 and OS-5 in a pSi sample comprised of a thin film of 9.3 $\mu$m thick. The experimental points obtained for the evolution of the stack filling fraction are also included in the same figure. 
The procedure to determine the time at which the thin film imbibition is completed and the imbibition of the PC starts ($\rm t_{\rm fill}$), was the following: first, $\rm t_{\rm fill}$ was determined for OS-5 by identifying the maximum of the second derivative of $\rm \lambda_0(t)$. This filling time was used to normalize the time axis for OS-5 experimental data ($\rm \lambda_0(t)$  $\rm \&$ $\rm \lambda_{\rm peak}(t)$). Then, the time axis of OS-2 experimental data was normalized with the $\rm t_{\rm fill}$ that best collapses $\rm \lambda_{\rm peak}(t)$ OS-2 and OS-5 curves. This method is accurate since the compared imbibition experiments were measured in different pieces of the same pSi sample, eliminating any possible morphological variation of the porous structures. It is relevant to note that the filling fraction corresponding to the thin film (indicated with a horizontal dotted line in Figure \ref{fig:Reff_fit}b) effectively occurs in coincidence with the arrival of the fluid front at the PC, which is evident by a slope change in $\rm \lambda_0(t)$. Moreover, the filling fraction of 0.8 matches the thickness measured through SEM images.

In Figure \ref{fig:Reff_fit}b, a fitting with the LWE (Eq. \ref{eq:LW}) has been included (solid line) to show that the capillary filling of both oligomers follows the classical $\rm \sqrt{t}$ dynamics in the uniform porosity thin film. The analysis continues by extracting $\rm r_{\rm eff}^{OS-2}$ and $\rm r_{\rm eff}^{OS-5}$ from Eq. \ref{eq:LW} with the corresponding $\rm t=t_{\rm fill}$ and $\rm L$ equal to the thin film thickness. 
The obtained results are summarized in Tab.~\ref{tab:results}. They indicate that  $\rm r_{\rm eff}^{OS-2}$ and $\rm r_{\rm eff}^{OS-5}$ are 5.8 and 11 times smaller than the $\rm r_{\rm eff}$ estimated for the glycerol imbibition, respectively. 
Using Eq. \ref{eq:r_eff_4}, the immobile layer thickness that is consistent with that change in $\rm r_{\rm eff}$ in each case can be estimated. The values obtained in this way are $\rm b_{\rm OS-2}=1.6$ nm and $\rm b_{\rm OS-5}=1.95$ nm. Quantum mechanical molecular size estimations resulted in ${\rm0.9-1.2~nm}$ for OS-2 and ${\rm1.4-1.8~nm}$ for OS-5.\cite{jensen2013molecule} The ratio of these sizes of 1.5 is comparable to the ratio of the immobile layer of 1.2. A way to interpret the immobile layer is to consider ${\rm 1-3}$ densely packed molecular layers\cite{Zhang2021}, with molecules stretched in the direction of drag and a slow diffusion at the wall. Note that the existence of such an hydrodynamically dead layer agrees with findings on other (simpler) molecular liquids \cite{Gruener2009, Vincent2016, Gruener2019}. 

It is interesting to compare these results with the tortuous pore model (Eq. \ref{eq:tau_L}), which requires $\tau$=6.3 for OS-2 and $\tau=$8.6 for OS-5 to predict the above effective radii. Beyond the extremely high values, $\tau$ results are different for each measurement, which is not compatible with the fact that the morphology of the thin film is the same in both cases.

\begin{table}
\caption{\label{tab:results} Results: The decelerated dynamics in capillary filling can be illustrated by apparent viscosities ${\rm \mu_{\rm app}}$, six and eleven fold higher than the bulk for OS-2 and OS-5, respectively. In the constriction model the different capillary rise dynamics of the oligomers is explained by the immobile layer thickness $b$. The squared fit parameter "a" of the main meniscus can be compared to the diffusion-like coefficient ${\rm D_{\rm PF}}$ of the precursor film spreading. Additionally, a thickness ${\rm d_{\rm PF}}$ for the precursor film is derived.}
\begin{ruledtabular}
\begin{tabular}{cccccc}
 &  ${\rm \mu_{\rm app}}$  [${\rm mPa~s}$] & ${\rm b}$ [${\rm nm}$] &${\rm a^2}$ [${\rm m^2/s}$] & ${\rm D_{\rm PF}}$ [${\rm m^2/s}$]&${\rm d_{\rm PF}}$ [${\rm nm}$]\\
\hline
OS-2 & 54  & 1.60& ${\rm 3.3{\cdot}10^{-10}}$& -& - \\
OS-5 & ${\rm 3.7{\cdot}10^{5}}$ & 1.95&${\rm 5.1{\cdot}10^{-14}}$& ${\rm 9.4{\cdot}10^{-13}}$& 0.9  \\

\end{tabular}
\end{ruledtabular}
\end{table}

\begin{figure}
\includegraphics[width=0.45\textwidth]{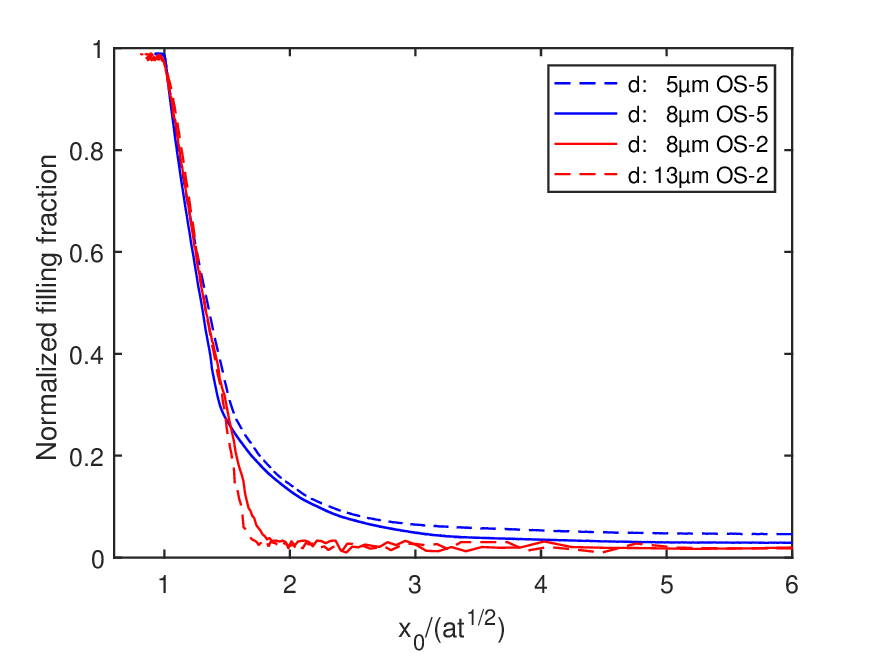}
\caption{\label{fig:menisci} Filling fraction of the photonic crystal with OS-2 and OS-5 versus a spatio-temporal rescaled variable. "a" is the prefactor of a square-root-of-time fit of the filling dynamics of the thin film in front of the PC. The filling time $t$ of the PC is rescaled with respect to its central position ${\rm x_0}$, which is adjusted for same starting positions of menisci of the respective oligomer. The thickness $d$ in the legend belongs to the pSi layer in front of the PC. It is measured by SEM.}
\end{figure}

Fig.~\ref{fig:menisci} displays the fluid front shapes of OS-2 and OS-5 for two samples each, with different porous layer thicknesses in front of their PC. Thinking in the chronological sequence in the experiment the figure must be read from right to left, since it has been re-scaled in space and time by ${\rm x_0/ (a \sqrt{t})}$. Here ${\rm x_0}$ stands for the middle of the PC and "a" is the measured coefficient for capillary rise in the preceding thin film. We calculated ${\rm x_0}$ for the abscissa position of one, by the point in time where the filling fraction is converging to a value of one. It can be noted, that the fluid front shape for the individual oligomers maintains their shape independent of the thin film layer thickness. OS-2 shows a sharp front whereas OS-5 has a broad slope, i.e. a precursor film, starting at a value ${\rm x_{\rm PF}}$ around 4.3 of the abscissa. The onset of the precursor film is determined by the intercept of a linear fit of the constant initial region and the tangent where the filling fraction starts to increase.\\ 
A diffusion like coefficient for the precursor film spreading can be derived by ${\rm D_{\rm PF}=(\tau~a~x_{\rm PF})^2 }$. For the experiments displayed in Fig.~\ref{fig:menisci} a mean of ${\rm 9.4\cdot10^{-13}~m^2/s}$ is obtained. Similar to MD simulations on nanopores\cite{Chibbaro2008a} the spreading dynamics of the precursor film follows a faster ${\rm \sqrt{t}}$-law than the main meniscus, as can be seen by comparing the squared prefactor of the capillary rise in Tab.~\ref{tab:results}. The value for the main meniscus of OS-5 is about twenty times smaller than its precursor film, which in turn is more than two orders of magnitude smaller than the capillary rise of OS-2. The latter explains why there is no precursor film observed for OS-2, even if one assumes a faster precursor film formation for this shorter oligomer. If one assumes that ${\rm D_{\rm PF}}$ scales similarly as a function of the degree of polymerization $N$ as in polymer surface diffusion, where Sukhishvili et al. found a proportionality of ${\rm D \sim N^{-3/2}}$, there is barely any increase in ${\rm D_{\rm PF}}$ expected for OS-2 compared to OS-5 \cite{sukhishvili2002surface}.  \\   
The end of the precursor film region is determined by deviation from a linear fit of the meniscus region by more than the standard deviation in 10 consecutive points. Taking the difference of the filling fraction limiting the precursor region, its maximal filling fraction ${\rm \Phi_{\rm PF}}$ can be assessed. A simple tube filling model starting from the pore wall leads to an approximation of the precursor thickness by ${\rm \Phi_{\rm PF}= \frac{2 \langle r_h \rangle d_{\rm PF} - d_{\rm PF}^2}{\langle r_h \rangle^2}}$, resulting to ${\rm d_{\rm PF} \approx 0.9~nm}$. \\

\section{conclusion}

The combined thin-film interference and photonic crystal analysis of pSi during liquid imbibition allows the detection of capillary rise and precursor film dynamics in the same experiment. Especially for polymeric liquids, the necessity to measure capillary rise directly becomes clear, as they exhibit different fluid dynamics when the pore size of the scaffold becomes comparable to the molecule size. In the experiments performed with OS-2 and OS-5, a strong slow-down was observed. This can be explained by an apparent viscosity of about six and eleven times the bulk value, respectively. We, however, explain this by the increasing importance of local undulations in the pore radius with the molecular size of the infiltrating liquid, but otherwise unchanged bulk fluid parameters. Indeed, a simplified constriction model with a dead zone consistently describes imbibition in the pSi thin films for both oligomers. This model is superior to the concept of a modified LWE with a tortuosity by having a consistent geometrical description. Moreover, the tortuosities were found to be unreasonably high, suggesting much more meandering pores than can be deduced from SEM images.\\

In the present study, the constriction model gives a very good description of the capillary rise dynamics. Even in the framework of the semi-continuum approach, the model relates the measured (macroscopic) variables to parameters that characterize the most important nanoscale features, namely the thickness of the hydrodynamic dead zone. Furthermore, the predicted values are in reasonable agreement with estimates of the molecular size of the oligomers.\\

Following the resonance wavelength of the PC, filling phenomena can be resolved with sub-nanometer precision. The coefficient of capillary rise dynamics determined in the same experiment is used to visualize the shape of the fluid front. For OS-2 the fluid front is flat, while for OS-5 a precursor film is detected. We attribute this to the fact that the pore surface diffusion dynamics in OS-5 are fast enough to allow a thin film to move beyond the main capillary front. In contrast, the fast imbibition rate of OS-2 precludes precursor film formation. We also show that our opto-fluidic technique allows to measure the maximum thickness and a surface diffusion-like coefficient of the OS-5 precursor film with remarkably high spatio-temporal resolution.\\ 

From an application point of view, the parameters obtained by the presented method could be used to rationally design polymer nanotubes by melt infiltration. Moreover, we envision that our experimental platform can be employed to determine the surface diffusion coefficients of polymers, which is generally a challenging task.\\

\begin{acknowledgements}

The authors acknowledge the funding support received from the following institutions: Deutsche Forschungsgemeinschaft (DFG, German Research Foundation) Project number 519853330, CRC 986 "Tailor-Made Multi-Scale Materials Systems" Project number 192346071, ”Dynamic Electrowetting at Nanoporous Surfaces: Switchable Spreading, Imbibition, and Elastocapillarity”, Project number 422879465 (SPP 2171), Consejo Nacional
de Investigaciones Cientificas y Tecnicas, grant PIP-2020-1049, Universidad Nacional del Litoral, grant
CAID 2020-50620190100114L.
\end{acknowledgements}

%
\end{document}